\documentclass[aps, prd, floats, floatfix, onecolumn, superscriptaddress, nofootinbib, showpacs, letterpaper, notitlepage, 10pt]{revtex4-1}

\usepackage{graphicx}
\usepackage{amsmath}
\usepackage{amsfonts}
\usepackage{amssymb}
\usepackage{amstext}
\usepackage{tensor}
\usepackage[top=1in, bottom=1in, left=1in, right=1in]{geometry}
\usepackage[english]{babel}
\usepackage{helvet}
\usepackage{microtype}
\usepackage[pdftex,
  pdftitle={A Simplified Approach to General Scalar-Tensor Theories},
  pdfauthor={Jolyon Bloomfield},
  bookmarks,bookmarksopen=false,
  pdfstartview={FitH},
  linktocpage=true
]{hyperref}

\def\d{{\rm d}}

\begin{document}

\title{A Simplified Approach to General Scalar-Tensor Theories}

\author{Jolyon Bloomfield}
\email{jkb84@cornell.edu}
\affiliation{Center for Radiophysics and Space Research, Cornell University, Ithaca, NY 14853}
\affiliation{Laboratory for Elementary Particle Physics, Cornell University, Ithaca, NY 14853}

\date{\today}

\begin{abstract}
    The most general covariant action describing gravity coupled to a scalar field with only second order equations of motion, Horndeski's theory (also known as ``Generalized Galileons''), provides an all-encompassing model in which single scalar dark energy models may be constrained. However, the generality of the model makes it cumbersome to manipulate. In this paper, we demonstrate that when considering linear perturbations about a Friedmann-Robertson-Walker background, the theory is completely specified by only six functions of time, two of which are constrained by the background evolution. We utilise the ideas of the Effective Field Theory of Inflation/Dark Energy to explicitly construct these six functions of time in terms of the free functions appearing in Horndeski's theory. These results are used to investigate the behavior of the theory in the quasistatic approximation. We find that only four functions of time are required to completely specify the linear behavior of the theory in this limit, which can further be reduced if the background evolution is fixed. This presents a significantly reduced parameter space from the original presentation of Horndeski's theory, giving hope to the possibility of constraining the parameter space. This work provides a cross-check for previous work on linear perturbations in this theory, and also generalizes it to include spatial curvature.
\end{abstract}

\pacs{04.50.Kd, 95.36.+x}

\maketitle

\section{Introduction}

Since the discovery of the accelerated expansion of the universe \cite{Riess1998,Perlmutter1999}, a plethora of models to explain this phenomena have been proposed (see Ref. \cite{Skordis2011} for a comprehensive review). In order to obtain a homogeneous and isotropic expansion, the mechanism behind the accelerated expansion is typically attributed to a (effective) scalar field. It is, however, generally straightforward to select parameters for a model to select the desired expansion history of the universe \cite{Huterer:1998qv,Starobinsky:1998fr,Nakamura:1998mt,Boisseau:2000pr,Park2010}, and so it is important to investigate the perturbative behavior of a model in order to establish observational limits on the parameter space. Unfortunately, this analysis is typically much harder to perform than the background analysis.

Recently, there has been much interest in general models that attempt to contain a number of individual models \cite{Bloomfield:2011wa,Gubitosi2012,Bloomfield2012,Baker2012,Silvestri2013,Charmousis2011,battye2012}. One of the motivations for these approaches is to be able to construct model-independent constraints, which will hopefully save time and effort analyzing individual models. This paper addresses the intersection of two such methods.

An old result by Horndeski \cite{Horndeski1974}, recently rediscovered by Deffayet \textit{et al.} \cite{Deffayet2011,Kobayashi2011}, states that the most general possible covariant scalar field theory coupled to gravity without higher order derivatives is given by the action
\begin{align}
  S = \sum_{i=2}^5 S_i = \sum_{i=2}^5 \int \d^4 x \sqrt{-g} \, {\cal L}_i \, ,
\end{align}
where
\begin{subequations}\label{eq:horndeski}
\begin{align}
{\cal L}_{2} &= K(\varphi, X),
\label{eqn:L2} \\
{\cal L}_{3} &= -G_{3}(\varphi, X) \Box\varphi,
\label{eqn:L3} \\
{\cal L}_{4} &= G_{4}(\varphi, X) R + G_{4,X} \, [(\Box\varphi)^2 - (\nabla_{\mu} \nabla_{\nu} \varphi) \, (\nabla^{\mu}\nabla^{\nu}\varphi)] ,
\label{eqn:L4} \\
{\cal L}_{5} &= G_{5}(\varphi,X)\, G_{\mu\nu}\,(\nabla^{\mu}\nabla^{\nu}\varphi)
\nonumber\\
&-\frac{1}{6}\, G_{5,X}\,[(\Box\varphi)^{3}-3(\Box\varphi)\,(\nabla_{\mu}\nabla_{\nu}\varphi)\, (\nabla^{\mu}\nabla^{\nu}\varphi)+2(\nabla^{\mu}\nabla_{\alpha}\varphi)\, (\nabla^{\alpha}\nabla_{\beta}\varphi)\,(\nabla^{\beta}\nabla_{\mu}\varphi)]\,.
\label{eqn:L5}
\end{align}
\end{subequations}
Here, $X = -g^{\mu\nu} \partial_\mu \varphi \partial_\nu \varphi / 2$\footnote{Note that there exist different definitions for $X$ in the literature, variously including/excluding the factor of 1/2 and the minus sign.}, and $K$, $G_3$, $G_4$ and $G_5$ are arbitrary free functions\footnote{We omit a matter action, which we take to be minimally coupled to the metric in the Jordan frame. This assumes that the weak equivalence principle holds.}. This action will henceforth be referred to as the ``Horndeski action''. The generality of this action suggests that almost all single scalar field dark energy models are contained within it (Lorentz violating models such as Ho\v{r}ava-Lifshitz gravity \cite{Horava2009} are the biggest category of exceptions). On the other hand, this description is almost too general, in that the impressive amount of freedom contained in the action prohibits the calculation of observational constraints for the model.

A particularly beautiful approach to perturbations in cosmological spacetimes comes from the Effective Field Theory (EFT) of Inflation \cite{Cheung2008}. In this approach, it is assumed that the background evolution of the universe is known, allowing the perturbative behavior to be investigated in detail. This approach has been generalized to dark energy models \cite{Creminelli2009,Gubitosi2012,Bloomfield2012}, where it has been demonstrated to provide a very general method for calculating perturbative behavior. By discarding information about the background, the EFT approach condenses the model-dependent information into a handful of functions of time. It is hoped that by placing model-independent constraints on these functions of time, we will be able to constrain very general theories of dark energy, such as Horndeski's theory.

The goal of this paper is to perform a linear perturbation analysis of Horndeski's theory about a Friedmann-Robertson-Walker (FRW) background by casting it into the formalism of the EFT approach. De Felice \textit{et al.} \cite{DeFelice2011} have already performed such a feat by calculating the linearized equations of motion and investigating the results. Here, we simplify matters somewhat by demonstrating that Horndeski's theory can be condensed from its original four functions of two variables down to just six functions of time, for the purposes of describing linear perturbations. Furthermore, we generalize upon the results of De Felice \textit{et al.} by performing the analysis with arbitrary spatial curvature. Once the results have been obtained, we use the machinery already developed for the EFT approach to calculate the modifications from general relativity in the quasistatic limit.

This paper is arranged as follows. We begin by providing a very brief overview of the EFT of Inflation formalism in Section \ref{sec:overview}. Next, we use counting arguments to decide which operators will be necessary to describe Horndeski's theory within the EFT framework in Section \ref{sec:operators}. We then demonstrate how the two approaches can be matched together, and explicitly detail the equivalence in Section \ref{sec:matching}. Finally, we investigate the perturbative behavior of the theory in the quasistatic approximation in Section \ref{sec:observables}, paying close attention to the number of free functions required for a complete description. We conclude by discussing the application of these results to inflationary models and suggest future directions.

\section{Overview of Effective Field Theory Approach}\label{sec:overview}
The EFT of Inflation formalism \cite{Cheung2008} provides a description of perturbations in single scalar field theories about an FRW background. It works by using the unitary gauge, in which the foliation of spacetime is chosen such that the scalar field has no perturbations (the scalar degree of freedom has been ``eaten'' by the metric). Because of the symmetries of FRW, the possible operators in an action in this gauge are very straightforward, and can be arranged into a perturbative expansion, with each operator having a function of time for a coefficient. In this section, we provide a very brief overview of the formalism, directing the reader to the literature for a thorough introduction \cite{Bloomfield2012,Cheung2008,Gubitosi2012}.

Begin by assuming that the background metric takes the form
\begin{align}
  \d s^2 = - \d t^2 + a^2 \tilde{g}_{ij} \d x^i \d x^j
\end{align}
where $\tilde{g}_{ij}$ is a maximally symmetric time-independent spatial metric with scalar curvature $R^{(3)} = 6k_0$. The operators that can be included in the unitary gauge action must be constructed of the following objects.
\begin{align}
  f(t), \quad
  \delta g^{00} = g^{00} + 1,
  \quad h_{ij},
  \quad \delta K_{ij} = K_{ij} - K_{ij}^0 ,
  \quad \delta \tensor*{R}{^{(3)}_{ijkl}} = \tensor*{R}{^{(3)}_{ijkl}} - \tensor*{R}{^{(3)0}_{ijkl}},
  \quad D_i,
  \quad \partial_t,
  \quad \epsilon^{ijk}
\label{eq:objects}
\end{align}
These are, in order, functions of time, a perturbation to the $g^{00}$ component of the inverse metric, the spatial metric, a perturbation to the extrinsic curvature tensor to surfaces of constant time\footnote{Different conventions exist for the sign of the extrinsic curvature tensor. We use the definition $K_{\mu \nu} = (\delta_\mu^\lambda + n_\mu n^\lambda) \nabla_\lambda n_\nu$, and define the normal to surfaces of constant time as $n_\mu = \partial_\mu t / \sqrt{- g^{00}}$.}, a perturbation to the Riemann tensor of the spatial metric, covariant derivatives associated with the spatial metric, time derivatives\footnote{Technically, the object that is invariant under the symmetries is $\partial_t - {\cal L}_{\vec{N}}$, where ${\cal L}_{\vec{N}}$ is the Lie derivative along the shift vector in an ADM decomposition. However, because the shift vector is vanishing on the FRW background, the terms involving the shift are higher order in perturbations and can be ignored.}, and the three-dimensional antisymmetric tensor. Evaluated on the background, the spatial metric  is given by $h_{ij} = a^2 \tilde{g}_{ij}$. Because there are no $\epsilon$ symbols appearing in Horndeski's theory, they will not be required in the perturbation description, and we ignore them from now on. The perturbations are constructed by subtracting the background values written in terms of just functions of time, the spatial metric and the normal vector.

It was previously shown \cite{Gubitosi2012,Bloomfield2012} that the leading order operators in such a construction are given by the following action.
\begin{align}
  S = \int \d^4x \sqrt{-g} {}& \bigg\{ \frac{m_0^2}{2} \Omega(t) R
  + \Lambda(t)
  - c(t) \delta g^{00}
  + \frac{M_2^4 (t)}{2} (\delta g^{00})^2
\nonumber \\
  & \qquad - \frac{\bar{M}_1^3 (t)}{2} \delta g^{00} \delta \tensor{K}{^i_i}
    - \frac{\bar{M}_2^2 (t)}{2} \delta \tensor{K}{^i_i^2}
    - \frac{\bar{M}_3^2 (t)}{2} \delta \tensor{K}{^i_j} \delta \tensor{K}{^j_i}
\nonumber \\
  & \qquad
    + \frac{\hat{M}^2 (t)}{2} \delta g^{00} \delta R^{(3)}
    + h^{ij} \partial_i \delta g^{00} \partial_j \delta g^{00}
    + \ldots
  \bigg\} + S_{matter}[g_{\mu \nu}] \label{eq:fullaction}
\end{align}
The functions $\Omega(t)$, $\Lambda(t)$ and $c(t)$ are related to the background evolution of the comology, and appear in the Friedmann equations as \cite{Gubitosi2012,Bloomfield2012}
\begin{align}
  3 m_0^2 \Omega \left[H^2 + \frac{k_0}{a^2} + H \frac{\dot{\Omega}}{\Omega} \right] &= \rho_m - \Lambda + 2 c\,, \label{eq:friedmann1}
\\
  m_0^2 \Omega \left[3 H^2 + 2 \dot{H} + \frac{k_0}{a^2} + \frac{\ddot{\Omega}}{\Omega} + 2 H \frac{\dot{\Omega}}{\Omega} \right] &= - \Lambda - P_m \label{eq:friedmann2}
\end{align}
where $\rho_m$ and $P_m$ are the energy density and pressure arising from matter and radiation fields, $H$ is the Hubble parameter $\dot{a}/a$, and overdots refer to time derivatives. Once the background evolution $H(t)$ as well as $\Omega(t)$ is specified, $c(t)$ and $\Lambda(t)$ also become completely specified, by virtue of these equations. This effectively reduces the number of free functions of time by one.

Through the use of a technique known as the ``St\"uckelberg trick'', the scalar field perturbations can be reintroduced to the action \eqref{eq:fullaction}. The perturbations are typically called $\pi$, which relate to the scalar field as
\begin{align}
  \varphi = \varphi^0(t + \pi) = \varphi^0 (t) + \delta \varphi
\end{align}
where $\varphi^0$ is the background scalar field. Both the equations of motion for the $\pi$ field and the stress-energy tensor were calculated in \cite{Bloomfield2012}.

\section{Choice of Operators}\label{sec:operators}
In order to construct linear perturbations in Horndeski's theory within the EFT formalism, we first need to identify which operators will be required. Using the notation in the action \eqref{eq:fullaction}, the operators with coefficients $\Omega(t)$, $\Lambda(t)$ and $c(t)$ will definitely be required, as these operators specify the background FRW evolution. The remaining operators, which only affect the behavior of the perturbations, should have the property that they contain only two derivatives\footnote{At least, at second order in perturbations. At higher orders, more terms can become involved in order to provide the required cancellations.}. Of course, we may take linear combinations of the remaining operators to cancel out any higher order derivatives.

For linear perturbation theory, we are interested in operators that are second order in perturbed quantities. Furthermore, all of these operators must be constructed out of the objects listed in Eq. \eqref{eq:objects}. Consider the metric in Newtonian gauge, which contains no derivatives acting on the scalar perturbations. The object $\delta g^{00}$ has no derivatives, nor does the spatial metric, while the Riemann tensor contains two spatial derivatives, and the extrinsic curvature tensor (being the Lie derivative of the spatial metric along the normal to hypersurfaces of constant time) contains one time derivative. Therefore, we can join these objects together to make a complete list of all possible operators at second order in perturbations that contain two derivatives or less\footnote{Gleyzes \textit{et al.} \cite{Gleyzes2013} point out that this condition is actually too restrictive; it is possible for a combination of operators containing higher derivatives to have their higher derivative terms cancel. They show that the combination $\tensor{R}{^{(3)}_i^j} \delta \tensor{K}{^i_j} - R^{(3)} \delta \tensor{K}{^i_i} / 2$ can be reduced to operators in this list however, and that the combination $2 (R^{(3)})^2 - 3/4 \tensor{R}{^{(3)}_i^j} \tensor{R}{^{(3)}_j^i}$ is a total derivative, even when multiplied by a function of time (note that in both these cases, the curvature scalar and tensor are represented without the background component subtracted off). We are unaware of further examples, and it is likely that in any case, such may be rewritten (as these may be) in terms of the operators presented here.}.
\begin{align}
  (\delta g^{00})^2,
  \quad \delta g^{00} \delta \tensor{K}{^i_i},
  \quad (\delta \tensor{K}{^i_i})^2,
  \quad \delta \tensor{K}{^i_j} \tensor{K}{^j_i},
  \quad \delta g^{00} \delta R^{(3)},
  \quad D_i \delta g^{00} D^i \delta g^{00},
  \quad (\partial_t \delta g^{00})^2,
  \quad \partial_t \delta g^{00} \delta \tensor{K}{^i_i} \label{eq:operators}
\end{align}
Here, indices have been raised and lowered using the spatial metric, and we have taken advantage of integration by parts.

Unfortunately, the ``two-derivative'' property is a little muddled, because once the St\"uckelberg trick has been applied, the number of derivatives acting on the St\"uckelberg field can be different from the rest of an operator. Here is how the St\"uckelberg field is introduced to each object, truncated at first order in perturbations.
\begin{align}
  \delta g^{00} & \rightarrow \delta g^{00} - 2 \dot{\pi}
\\
  \delta \tensor{K}{^i_j} & \rightarrow \delta \tensor{K}{^i_j} + \dot{H} \pi \delta^i_j
  + D^i D_j \pi
\\
  \delta R^{(3)} & \rightarrow \delta R^{(3)} + 4 H D^2 \pi + 12 H \frac{k_0}{a^2} \pi
\end{align}
In Table \ref{tab:derivatives}, we count the number and type of derivatives acting on the metric and scalar field perturbations for each of the operators in \eqref{eq:operators}.

\begin{table}[b]
  \centering
  \noindent\makebox[\textwidth]{%
  \begin{tabular}{|c||c|c|c|}
    \hline
    \textbf{Operator} &
    Metric-Metric &
    Metric-Scalar &
    Scalar-Scalar
    \\ \hline\hline
    $(\delta g^{00})^2$ & - & 1t & 2t \\ \hline
    $\delta g^{00} \delta \tensor{K}{^i_i}$ & 1t & 2t + 2s & 2s1t \\ \hline
    $(\delta \tensor{K}{^i_i})^2$ & 2t & 2s1t & 4s \\ \hline
    $\delta \tensor{K}{^i_j} \tensor{K}{^j_i}$ & 2t & 2s1t & 4s \\ \hline
    $\delta g^{00} \delta R^{(3)}$ & 2s & 2s1t & 2s1t \\ \hline
    $D^i \delta g^{00} D_i \delta g^{00}$ & 2s & 2s1t & 2s2t \\ \hline
    $(\partial_t \delta g^{00})^2$ & 2t & 3t & 4t \\ \hline
    $\partial_t \delta g^{00} \delta \tensor{K}{^i_i}$ & 2t & 3t + 2s1t & 2s2t \\ \hline
  \end{tabular}}
  \caption[Number of space and time derivatives in operators]{
    The number of space and time derivatives in different operators, sorted by whether the derivatives are acting on metric perturbations only, scalar perturbations only, or a mix of the two. Only the highest order derivatives in each category are listed. \\
    \begin{tabular}{p{2cm}cp{0.7 \textwidth}}
      &- & No derivatives are present\\
      &$n$s & There are $n$ spatial derivatives\\
      &$n$t & There are $n$ time derivatives\\
      &+ & Multiple combinations of derivatives are present at the highest order
    \end{tabular}
  }
\label{tab:derivatives}
\end{table}

We see from the table that the operator $(\delta g^{00})^2$ is fine, as only second order derivatives are present. The operator $\delta g^{00} \delta \tensor{K}{^i_i}$ might be problematic because of the cross term $- 2 \dot{\pi} D^2 \pi$, but the time derivative can be removed using integration by parts, and so this operator is also fine.

The operator $(\partial_t \delta g^{00})^2$ is alone in that it contains four time derivatives acting on scalar perturbations, so these derivatives cannot be cancelled in combination with another operator, and this operator must be dropped. Similarly, the operator $\partial_t \delta g^{00} \delta K$ contains terms with three time derivatives, which cannot be cancelled with anything else, and so this operator must also be discounted. The operator $D_i \delta g^{00} D^i \delta g^{00}$ contains terms with two time derivatives and two spatial derivatives which cannot be cancelled in combination with any remaining terms, and so this term is also thrown out.

The operators $(\delta K)^2$ and $\delta \tensor{K}{^i_j} \delta \tensor{K}{^j_i}$ both contain terms with four spatial derivatives acting on the scalar perturbations. A simple analysis reveals that so long as the coefficients of these operators are equal but negative of each other, these terms cancel. Both operators also contain terms involving one time and two spatial derivatives. Looking at the table, it is feasible for the operator $\delta g^{00} \delta R^{(3)}$ to provide an exact cancellation for this term. Indeed, these terms do cancel when the coefficient of $\delta g^{00} \delta R^{(3)}$ is given by
\begin{align}
  2 \hat{M}^2 = \bar{M}_2^2 = - \bar{M}_3^2.
\end{align}
To see this cancellation at the quadratic order requires expanding the three operators in metric perturbations and using integration by parts, which we demonstrate in the following section. The operator $\delta g^{00} \delta R^{(3)}$ also has a cross term $\dot{\pi} D^2 \pi$ which can be dealt with through integration by parts, just as for $\delta g^{00} \delta \tensor{K}{^i_i}$.

Thus, we expect that linear perturbations in Horndeski's general theory about an FRW background will be described in the EFT of Dark Energy context by the following action.
\begin{align}
  S = \int \d^4x \sqrt{-g} {}& \bigg\{ \frac{m_0^2}{2} \Omega(t) R
  + \Lambda(t)
  - c(t) \delta g^{00}
  + \frac{M_2^4 (t)}{2} (\delta g^{00})^2
\nonumber \\
  & \qquad - \frac{\bar{M}_1^3 (t)}{2} \delta g^{00} \delta \tensor{K}{^i_i}
    - \frac{\bar{M}_2^2 (t)}{2} \left(\delta \tensor{K}{^i_i^2}
      - \delta \tensor{K}{^i_j} \delta \tensor{K}{^j_i}
      + 2 \delta g^{00} \delta R^{(3)} \right)
  \bigg\} + S_{matter}[g_{\mu \nu}] \label{eq:guessedaction}
\end{align}
We claimed this result without proof in \cite{Bloomfield2012}. All that now remains is to identify the correspondence between the six functions of time appearing in this action and the free functions appearing in the Horndeski action.

\section{Matching Horndeski to the EFT Construction}\label{sec:matching}
The Horndeski action is sufficiently complicated that performing explicit matchings between the various functions by hand is a daunting task. Instead, we pursue a different approach, turning to computer algebra software.

As we are considering linear perturbations about a background in Horndeski's theory, we need to consider quantities that are quadratic in perturbations in the action. It is thus sufficient to expand the action for Horndeski's theory to quadratic order about a background, and compare this to the EFT action expanded to the same order. By matching coefficients in the action, we can thus express the functions of time in the EFT formalism in terms of background quantities in Horndeski's theory. Note that the zeroth order term in this expansion just gives the background value of the action, which is a meaningless boundary term for our purposes. The first order terms yield the background equations of motion, and thus must agree, as must the second order terms, which describe the equations of motion for the perturbations.

We work in unitary gauge in which there are no scalar field perturbations, and so the scalar field is given by $\varphi = \varphi^0(t)$ (the superscript zero will be dropped from now on). The subgroup of diffeomorphisms that preserve this gauge condition contains one free scalar function (as well as two functions associated with a divergenceless vector), as time reparameterization has been used to obtain the unitary gauge. Thus, only one of the four scalar components of the metric can be gauged away, and we need to consider a metric with three scalar perturbations, which we write as
\begin{align}
  \d s^2 = - (1 + 2 \phi) \d t^2 + 2 \partial_i B \d x^i \d t + a^2 \tilde{g}_{ij} (1 - 2 \psi) \d x^i \d x^j. \label{eq:unitarymetric}
\end{align}
The three scalar perturbations are $\phi$, $\psi$ and $B$.

When we expand the action in terms of these perturbations, there will be a number of different terms that are quadratic in the perturbations. However, some of these can be manipulated using integration by parts. We will preferentially integrate by parts to move spatial derivatives onto $B$ where possible, then $\psi$. Similarly, we will move all time derivatives onto $\psi$.

\subsection{Quadratic Expansion: EFT}
The EFT action we wish to expand is
\begin{align}
  S = \int \d^4x \sqrt{-g} {}& \bigg\{ \frac{m_0^2}{2} \Omega(t) R
  + \Lambda(t)
  - c(t) \delta g^{00}
  + \frac{M_2^4 (t)}{2} (\delta g^{00})^2
\nonumber \\
  & \qquad - \frac{\bar{M}_1^3 (t)}{2} \delta g^{00} \delta \tensor{K}{^\mu_\mu}
    - \frac{\bar{M}_2^2 (t)}{2} \delta \tensor{K}{^\mu_\mu^2}
    - \frac{\bar{M}_3^2 (t)}{2} \delta \tensor{K}{^i_j} \delta \tensor{K}{^j_i}
    + \frac{\hat{M}^2 (t)}{2} \delta g^{00} \delta R^{(3)}
  \bigg\},
\end{align}
where we have left the last three functions of time separate in order to demonstrate the required cancellations of higher derivative terms. Using the metric \eqref{eq:unitarymetric}, it is straightforward to express the perturbed objects in this action to the appropriate order.
\begin{align}
  \sqrt{-g} &= a^3 \left( 1 + \phi - 3 \psi - \frac{\phi^2}{2} - 3 \phi \psi + \frac{3}{2} \psi^2 + \frac{1}{2} \frac{\tilde{g}^{ij}}{a^2} \partial_i B \partial_j B \right)
\\
  \delta g^{00} &= 2 \phi - 4 \phi^2 + \frac{\tilde{g}^{ij}}{a^2} \partial_i B \, \partial_j B
\\
  \delta \tensor{K}{^i_j} &= \delta^i_j (H \phi + \partial_t \psi) + \frac{\tilde{g}^{ik}}{a^2} D_k D_j B
\\
  \delta R^{(3)} &= \frac{12 k_0 \psi + 4 \tilde{g}^{ij} D_i D_j \psi}{a^2}
\end{align}
The expansion of the Ricci scalar, including the metric determinant and integrating by parts, is a straightforward but somewhat tedious calculation.
\begin{align}
  S_{R} = \int \d^4 x \, a^3 \frac{m_0^2}{2} \bigg\{ &
  \Omega R^0
  + 6 \left( H^2 \Omega + H \dot{\Omega} + \Omega \frac{k_0}{a^2} \right) \phi
  - 6 \left( 3 H^2 \Omega + \Omega \frac{k_0}{a^2} + 2 \Omega \dot{H} + 2 H \dot{\Omega} + \ddot{\Omega} \right) \psi
\nonumber \\ &
  - 3 \left(3 H^2 \Omega + 3 H \dot{\Omega} + \Omega \frac{k_0}{a^2}\right) \phi^2
  - 2 \Omega \psi \tilde{\square} \psi
  + 3 \left(3 H^2 \Omega + 2 \Omega \dot{H} + 2 H \dot{\Omega} + \ddot{\Omega} - \Omega\frac{k_0}{a^2}\right) \psi^2
\nonumber \\ &
  + 4 \Omega \phi \tilde{\square} \psi
  - 6 \left(3 H^2 \Omega + 3 H \dot{\Omega} + \Omega \frac{k_0}{a^2}\right) \phi \psi
  - \left(3 H^2 \Omega + 3 H \dot{\Omega} + \Omega \frac{k_0}{a^2}\right) B \tilde{\square} B
\nonumber \\ &
  - 4 \Omega \dot{\psi} \tilde{\square} B
  - 6 \Omega \dot{\psi}^2
  - 2 (2 H \Omega + \dot{\Omega}) \phi \tilde{\square} B - 6 (2 H \Omega + \dot{\Omega}) \phi \dot{\psi}
  \bigg\}
\end{align}
Here, $R^0 = 6 \dot{H} + 12 H^2 + 6 k_0/a^2$ is the background Ricci scalar, and $\tilde{\square} = (1/a^2) \tilde{g}^{ij} D_i D_j$.

Combining all of the terms in the action yields the following expression for the linear and quadratic terms.
\begin{align}
  S_{EFT} = \int \d^4 x \, a^3 \bigg\{ &
  \left[3 m_0^2 \left( H^2 \Omega + H \dot{\Omega} + \Omega \frac{k_0}{a^2} \right) - 2c + \Lambda \right] \phi
  - 3 \left[ m_0^2 \left( 3 H^2 \Omega + \Omega \frac{k_0}{a^2} + 2 \Omega \dot{H} + 2 H \dot{\Omega} + \ddot{\Omega} \right) + \Lambda \right] \psi
\nonumber \\ &
  - \left[\frac{3}{2} m_0^2 \left(3 H^2 \Omega + 3 H \dot{\Omega} + \Omega \frac{k_0}{a^2}\right) - 2 c + \frac{\Lambda}{2} - 2 M_2^4 + 3 H \bar{M}_1^3 + \frac{9}{2} H^2 \bar{M}_2^2 + \frac{3}{2} \bar{M}_3^2 H^2 \right] \phi^2
\nonumber \\ &
  - m_0^2 \Omega \psi \tilde{\square} \psi
  + \left[\frac{3}{2} m_0^2 \left(3 H^2 \Omega + 2 \Omega \dot{H} + 2 H \dot{\Omega} + \ddot{\Omega} - \Omega\frac{k_0}{a^2}\right) + \frac{3}{2} \Lambda \right]\psi^2
\nonumber \\ &
  + \left[ 2 m_0^2 \Omega + 4 \hat{M}^2 \right] \phi \tilde{\square} \psi
  + \left[- 3 m_0^2 \left(3 H^2 \Omega + 3 H \dot{\Omega} + \Omega \frac{k_0}{a^2}\right) + 6 c - 3 \Lambda + 12 \hat{M}^2 \frac{k_0}{a^2} \right]\phi \psi
\nonumber \\ &
  - \left[\frac{m_0^2}{2} \left(3 H^2 \Omega + 3 H \dot{\Omega} + \Omega \frac{k_0}{a^2}\right) + \frac{\Lambda}{2} - c + \bar{M}_3^2 \frac{k_0}{a^2} \right] B \tilde{\square} B
\nonumber \\ &
  - \left[2 m_0^2 \Omega + 3 \bar{M}_2^2 + \bar{M}_3^2 \right] \dot{\psi} \left(\tilde{\square} B + \frac{3}{2} \dot{\psi} \right)
\nonumber \\ &
  - \left[m_0^2 (2 H \Omega + \dot{\Omega}) + \bar{M}_1^3 + 3 H \bar{M}_2^2 + H \bar{M}_3^2 \right] \phi (\tilde{\square} B + 3 \dot{\psi})
  - \frac{1}{2} \left[ \bar{M}_2^2 + \bar{M}_3^2 \right] \tilde{\square} B \tilde{\square} B
  \bigg\} \label{eq:EFTquadratic}
\end{align}
Note that although there are thirteen coefficients to match, only eleven of these are independent. As we have eight functions of time in the action, this leaves three extra equations to ensure consistency.

Two operators in this action look problematic from the perspective of derivative counting. The most obvious is the last, $\tilde{\square} B \tilde{\square} B$, which evidently has too many spatial derivatives, enforcing $\bar{M}_2^2 = - \bar{M}_3^2$. The other is $\dot{\psi} \tilde{\square} B$. Here, the number of derivatives present is not completely obvious, and the issue is clouded by gauge. When we vary the action with respect to $B$, each term in the equation of motion is multiplied by $k^2$, and it appears that the equation of motion contains a third order derivative. However, if we write the shift vector $g_{i0} = \tilde{\nabla}_i B$, then $\dot{\psi} \tilde{\square} B$ contains only two derivatives of the metric, and varying with respect to the shift instead of $B$ then suggests that one of those two powers of $k$ comes from the parametrization of the shift. Indeed, Gleyzes \textit{et al.} \cite{Gleyzes2013} show that by solving the constraint equations and integrating them out of the action, the resulting action is strictly second order in derivatives, and so this operator is perfectly well-behaved. On the other hand, once the St\"uckelberg trick has been performed (changing $B \rightarrow \pi$, $\phi \rightarrow \phi - \dot{\pi}$, $\psi \rightarrow \psi - H \pi$, in Newtonian gauge) the term $\phi \tilde{\square} \psi$ yields a term $\dot{\pi} \tilde{\square} \psi$, which combines with $\dot{\psi} \tilde{\square} B \rightarrow \dot{\psi} \tilde{\square} \pi$, and appears to not have any confusion relating to the shift. The requirement that this three-derivative operator cancels yields
\begin{align}\label{eq:extra}
  4\hat{M}^2 - 3 \bar{M}_2^2 - \bar{M}_3^2 = 0,
\end{align}
giving the result $2 \hat{M}^2 = \bar{M}_2^2$ stated previously. This difference between the scalar field and unitary gauge arises from demanding that the physical degree of freedom have second order equations of motion, compared to demanding that the action be explicitly second order in derivatives. When we perform the matching to Horndeski's theory, the relationship \eqref{eq:extra} is found to be satisfied, despite the apparently good behavior of the physical degree of freedom when it is not. Thus, we expect that beyond linear perturbations, higher derivatives will appear if this relationship is not satisfied.

\subsection{Quadratic Expansion: Horndeski}
We investigate each of the terms in the Horndeski action in turn to demonstrate how they can be written in terms of operators in the EFT formalism. Note that the contributions from different terms to the coefficients of each operator in the EFT add linearly, so it is sufficient to investigate them in isolation.

\vspace{2mm}
$\mathbf{L_2}$\\
\indent We begin with the term ${\cal L}_{2} = K(\varphi, X)$. The mapping of this operator to the EFT construction has been demonstrated in the context of \textit{k}-essence models in a variety of papers, including \cite{Cheung2008, Creminelli2009, Gubitosi2012}. Because the construction is trivial, we do not use the machinery from above.

In unitary gauge, $X = (1 - \delta g^{00}) X_0$, with $X_0 = \dot{\varphi}^2 / 2$. Therefore, we simply expand $K$ in powers of $\delta g^{00}$, using the chain rule as necessary.
\begin{align}
  K(\phi, X) = K(\phi, X_0) - K_{,X} X_0 \delta g^{00} + \frac{1}{2} K_{,XX} X_0^2 (\delta g^{00})^2 + \ldots
\end{align}
Here, $K_{,X}$ indicates a derivative of $K$ with respect to $X$ (such derivatives are always evaluated on the background $(\phi, X_0)$). We can then read off the contribution to the following coefficients, with all others vanishing.
\begin{align}
  \Lambda(t) = K(\phi, X_0) \,,
\qquad
  c(t) = \frac{1}{2} K_{,X} \dot{\varphi}^2 \,,
\qquad
  M_2^4 = \frac{1}{4} K_{,XX} \dot{\varphi}^4.
\end{align}

\vspace{2mm}
$\mathbf{L_3}$\\
\indent We next look at the term ${\cal L}_{3} = - G_{3}(\varphi, X) \Box\varphi$. This term has been previously cast in the EFT language in \cite{Gubitosi2012}, although only results under a simplifying assumption were presented. The expansion of this term (and the following two) into the form of \eqref{eq:EFTquadratic} is rather long and uninsightful, so we present only the coefficients that we have extracted by matching the results with the EFT expansion.
\begin{align}
  c(t) &= \frac{1}{2} \dot{\varphi} \left(G_{3,X} (3 H \dot{\varphi}^2 + \dot{\varphi} \ddot{\varphi}) - 2 \dot{G}_3 \right)
\\
  \Lambda(t) &= - \dot{G}_{3} \dot{\varphi}
\\
  M_2^4 (t) &= \frac{1}{4} \dot{\varphi}^2 \left(G_{3,X} (3 H \dot{\varphi} + \ddot{\varphi}) + G_{3,XX} \dot{\varphi}^2 (3 H \dot{\varphi} + \ddot{\varphi}) - \dot{G}_{3,X} \dot{\varphi}\right)
\\
  \bar{M}_1^3 (t) &= - G_{3,X} \dot{\varphi}^3
\\
  \bar{M}_2^2 (t) &=
  - \bar{M}_3^2 (t) =
  2 \hat{M}^2 (t) =
  \Omega(t) = 0
\end{align}
Note that we choose to eliminate partial derivatives with respect to $\varphi$ in favour of time derivatives, where we use the chain rule
\begin{align}
  \dot{G}_3 = G_{3,\varphi} \dot{\varphi} + G_{3,X} \dot{\varphi} \ddot{\varphi}.
\end{align}

\vspace{2mm}
$\mathbf{L_4}$\\
\indent This is the first term that requires all operators to be present.
\begin{align}
  m_0^2 \Omega(t) &= 2 G_4
\\
  c(t) &= G_{4,X} \left[ \dot{\varphi}^2 \left( 3 H^2 - 2 \dot{H} + 3 \frac{k_0}{a^2} \right) - \ddot{\varphi}^2 - \dot{\varphi}(H \ddot{\varphi} + \dddot{\varphi}) \right] + 3 H G_{4,XX} \dot{\varphi}^3 (H \dot{\varphi} + \ddot{\varphi}) - \dot{G}_{4,X} (5 H \dot{\varphi}^2 + \dot{\varphi} \ddot{\varphi})
\\
  \Lambda(t) &= -2 G_{4,X} ( (3 H^2 + 2 \dot{H}) \dot{\varphi}^2 + \ddot{\varphi}^2 + 4 H \dot{\varphi} \ddot{\varphi} + \dot{\varphi} \dddot{\varphi} ) - 2 \dot{G}_{4,X} (2 H \dot{\varphi}^2 + \dot{\varphi} \ddot{\varphi})
\\
  M_2^4 (t) &=
  \frac{1}{2} G_{4,X} (2 \dot{H} \dot{\varphi}^2 + \ddot{\varphi}^2 + H \dot{\varphi} \ddot{\varphi} + \dot{\varphi} \dddot{\varphi})
  + \frac{3}{2} G_{4,XX} \dot{\varphi}^3 \left( 3 H^2 \dot{\varphi} + 2 H \ddot{\varphi} + \frac{k_0}{a^2} \dot{\varphi} \right)
\nonumber \\ & \qquad
  + \frac{3H}{2} G_{4,XXX} \dot{\varphi}^5 (H \dot{\varphi} + \ddot{\varphi})
  + \frac{1}{2} \dot{G}_{4,X} (\dot{\varphi} \ddot{\varphi} - 4 H \dot{\varphi}^2)
  - \frac{3H}{2} \dot{\varphi}^4 \dot{G}_{4,XX}
\\
  \bar{M}_1^3 (t) &= -2 G_{4,X} (2 H \dot{\varphi}^2 + \dot{\varphi} \ddot{\varphi}) - 2 \dot{\varphi}^3 G_{4,XX} (2 H \dot{\varphi} + \ddot{\varphi}) + 2 \dot{G}_{4,X} \dot{\varphi}^2
\\
  \bar{M}_2^2 (t) &=
  - \bar{M}_3^2 (t) =
  2 \hat{M}^2 (t) = - 2 G_{4,X} \dot{\varphi}^2
\end{align}

\clearpage
$\mathbf{L_5}$
This second term requires the same operators to be present as the previous one; the difference between these two terms only becomes evident at nonlinear order.
\begin{align}
  m_0^2 \Omega(t) &= - \dot{\varphi} \dot{G}_5
\\
  c(t) &=
  \frac{1}{2} G_{5,X} \left( \left[ 3 H^2 - 2 \dot{H} + 3 \frac{k_0}{a^2} \right] H \dot{\varphi}^3 - 4 H \dot{\varphi} \ddot{\varphi}^2 + \left[ 3 H^2 - 2 \dot{H} + 3 \frac{k_0}{a^2} \right] \dot{\varphi}^2 \ddot{\varphi} - 2 H \dot{\varphi}^2 \dddot{\varphi} \right)
\nonumber \\ & \qquad
  + \frac{H^2}{2} G_{5,XX} \dot{\varphi}^4 (H \dot{\varphi} + 3 \ddot{\varphi})
  + \frac{1}{2} \dot{G}_5 \left[ \left( - 6 H^2 + 4 \dot{H} - 6 \frac{k_0}{a^2} \right) \dot{\varphi} + H \ddot{\varphi} + \dddot{\varphi} \right]
\nonumber \\ & \qquad
  - H \dot{G}_{5,X} \dot{\varphi}^2 (2 H \dot{\varphi} + \ddot{\varphi})
  + \ddot{G}_5 \left( \frac{H}{2} \dot{\varphi} + \ddot{\varphi} \right)
  + \frac{1}{2} \dddot{G}_5 \dot{\varphi}
\\
  \Lambda(t) &=
  - 2 G_{5,X} \left( (H^2 + \dot{H}) H \dot{\varphi}^3 + 2 H \dot{\varphi} \ddot{\varphi}^2 + (3 H^2 + \dot{H}) \dot{\varphi}^2 \ddot{\varphi} + H \dot{\varphi}^2 \dddot{\varphi} \right)
  - H \dot{G}_{5,X} \dot{\varphi}^2 (H \dot{\varphi} + 2 \ddot{\varphi})
\nonumber \\ & \qquad
  + \dot{G}_5 \left( ( 6 H^2 + 4 \dot{H}) \dot{\varphi} + 4 H \ddot{\varphi} + \dddot{\varphi}\right)
  + 2 \ddot{G}_5 (2 H \dot{\varphi} + \ddot{\varphi})
  + \dddot{G}_5 \dot{\varphi}
\\
  M_2^4 (t) &=
  \frac{1}{4} G_{5,X} \left( \left[ 3 H^2 + 2 \dot{H} + 3 \frac{k_0}{a^2} \right] H \dot{\varphi}^3 + 4 H \dot{\varphi} \ddot{\varphi}^2 + \left[ 3 H^2 + 2 \dot{H} + 3 \frac{k_0}{a^2} \right] \dot{\varphi}^2 \ddot{\varphi} + 2 H \dot{\varphi}^2 \dddot{\varphi} \right)
\nonumber \\ & \qquad
  + \frac{3}{4} G_{5,XX} \dot{\varphi}^4 \left( \left[ 2 H^2 + \frac{k_0}{a^2} \right] H \dot{\varphi} + \left[ 4 H^2 + \frac{k_0}{a^2} \right] \ddot{\varphi} \right)
  + \frac{1}{4} G_{5,XXX} H^2 \dot{\varphi}^6 (H \dot{\varphi} + 3 \ddot{\varphi})
\nonumber \\ & \qquad
  - \frac{1}{4} \dot{G}_5 (4 \dot{H} \dot{\varphi} + H \ddot{\varphi} + \dddot{\varphi})
  - \frac{1}{4} \ddot{G}_5 (H \dot{\varphi} + 2 \ddot{\varphi})
  - \frac{1}{4} \dddot{G}_5 \dot{\varphi}
\nonumber \\ & \qquad
  + \frac{1}{4} \dot{G}_{5,X} \dot{\varphi}^2 \left( - \left[ 11 H^2 + 3 \frac{k_0}{a^2} \right] \dot{\varphi} + 2 H \ddot{\varphi} \right)
  - \frac{3}{4} \dot{G}_{5,XX} H^2 \dot{\varphi}^5
\\
  \bar{M}_1^3 (t) &=
  - G_{5,X} \dot{\varphi}^2 \left( \left[ 3 H^2 + \frac{k_0}{a^2} \right] \dot{\varphi} + 4 H \ddot{\varphi} \right)
  - H G_{5,XX} \dot{\varphi}^4 (H \dot{\varphi} + 2 \ddot{\varphi})
  + \dot{G}_5 (4 H \dot{\varphi} + \ddot{\varphi})
  + \ddot{G}_5 \dot{\varphi}
  + 2 H \dot{G}_{5,X} \dot{\varphi}^3
\\
  \bar{M}_2^2 (t) &=
  - \bar{M}_3^2 (t) =
  2 \hat{M}^2 (t) = - G_{5,X} \dot{\varphi}^2 (H \dot{\varphi} + \ddot{\varphi}) + 2 \dot{G}_5 \dot{\varphi}
\end{align}

\subsection{Summary}
We have demonstrated that a combination of EFT operators, chosen such that at most second order derivatives exist in the action, is capable of describing linear perturbations to the Horndeski action about an FRW background. The matching was performed very generally, including spatial curvature and an arbitrary background scalar field $\varphi(t)$, subject to the condition $|\dot{\varphi}| > 0$\footnote{Note that the unitary gauge cannot be defined piecewise, avoiding points at which $\dot{\varphi} = 0$, as the background scalar $\varphi$ is being used as a clock, and as such, must be a good time coordinate.}. The EFT operators required to match each term in the Horndeski action are summarised in Table \ref{tab:matching}.

\begin{table}[t]
  \centering
  \noindent\makebox[\textwidth]{%
  \begin{tabular}{|c||c|c|c|c|c|c|}
    \hline
    \textbf{Term} &
    $\Omega$ &
    $\Lambda$ &
    $c$ &
    $M_2^4$ &
    $\bar{M}_1^3$ &
    $\bar{M}_2^2 = - \bar{M}_3^2 = 2 \hat{M}^2$
    \\ \hline\hline
    ${\cal L}_2$ &  & \checkmark & \checkmark & \checkmark &  &  \\ \hline
    ${\cal L}_3$ &  & \checkmark & \checkmark & \checkmark & \checkmark &  \\ \hline
    ${\cal L}_4$ & \checkmark & \checkmark & \checkmark & \checkmark & \checkmark & \checkmark \\ \hline
    ${\cal L}_5$ & \checkmark & \checkmark & \checkmark & \checkmark & \checkmark & \checkmark \\ \hline
  \end{tabular}}
  \caption[EFT operators used in matching to Horndeski]{
    The EFT operators needed to match each term in the Horndeski action.
  }
\label{tab:matching}
\end{table}

Although we have only performed the matching using scalar perturbations, given that there are no other operators with at most two derivatives, the matching must also hold for tensor and vector perturbations. Thanks to the decomposition theorem, the scalar, vector and tensor modes are decoupled at linear order.

\section{Observables}\label{sec:observables}
Given the action \eqref{eq:guessedaction}, we may proceed to transform out of unitary gauge, and calculate the equations of motion. The procedure for the transformation is detailed in Refs. \cite{Cheung2008,Bloomfield2012,Gubitosi2012}, while the resulting scalar equations of motion are presented in \cite{Bloomfield2012}. Here, we use these results to derive the modified Poisson and anisotropic shear stress equations, as well as the scalar field equations of motion.

We work in Newtonian gauge, using the metric
\begin{align}
  \d s^2 = - (1 + 2 \phi) \d t^2 + a^2 \tilde{g}_{ij} (1 - 2 \psi) \d x^i \d x^j.
\end{align}
We follow the conventions of Kodama and Sasaki \cite{kodama1984} for the spatial mode functions in curved space. For each $\vec{k}$, define $Y_{\vec{k}}(x^i)$ to be a solution of the equation
\begin{align}
  \tilde{g}^{ij} D_i D_j Y_{\vec{k}} = - k^2 Y_{\vec{k}}\,,
\end{align}
where the spatial covariant derivatives are now those associated with the spatial metric $\tilde{g}_{ij}$ (these derivatives are the same as for the spatial metric $h_{ij}$ previously, when evaluated on the background). We will typically suppress the $\vec{k}$ dependence of $Y$. Both $Y$ and $Y^*$ will be solutions of this equation, as will any linear combination. We fix this freedom by choosing
\begin{align}
  \lim_{k_0 \rightarrow 0} Y(x^i) =  e^{i \vec{k} \cdot \vec{x}}\,,
\end{align}
so that the modes become the usual Fourier modes when the background spatial metric is flat. The normalization of $Y$ is chosen such that
\begin{align}
  \int d^3 x \sqrt{\tilde{g}} Y_{\vec{k}} Y^*_{\vec{k}^\prime} = (2 \pi)^3 \delta^3(\vec{k} - \vec{k}^\prime).
\end{align}
Taking derivatives of $Y$, vector and tensor mode functions are defined as
\begin{align}
  Y_i &= - k^{-1} D_i Y\,,
\\
  Y_{ij} &= k^{-2} D_i D_j Y + \frac{1}{3} \tilde{g}_{ij} Y.
\end{align}
We raise and lower indices on $Y_i$ and $Y_{ij}$ with $\tilde{g}_{ij}$ only. This ensures that $Y$, $Y_i$, and $Y_{ij}$ are independent of time, no matter the position of their indices.

We use these mode functions to decompose the matter stress-energy tensor in momentum space as
\begin{align}
  \tensor{T}{^0_0} &= - (\rho_m + \delta \rho Y)
\\
  \tensor{T}{^0_i} &= (\rho_m + P_m) v Y_i
\\
  \tensor{T}{^i_j} &= \left(P_m + \delta P Y \right) \delta^i_j + P_m \Pi \tensor{Y}{^i_j}.
\end{align}
We define
\begin{align}
  \rho_m \Delta = \delta \rho
  + \frac{3H}{k} (\rho_m + P_m) v
\end{align}
as the rest-frame density perturbation of matter.

Using the results of \cite{Bloomfield2012}, the modified Poisson equation arising from the action \eqref{eq:guessedaction} is given by
\begin{align}
  m_0^2 \Omega \frac{2k^2 - 6k_0}{a^2} \psi
  = {}&
  - \rho_m \Delta
  - 2 c (\dot{\pi} - \phi)
  + m_0^2 \dot{\Omega} \left[
  - 3 \dot{H} \pi
  + \frac{k^2}{a^2} \pi
  - 3 \left(\dot{\psi} + H \phi \right) \right]
  - 4 M_2^4 (\dot{\pi} - \phi)
\nonumber \\ &
  - \bar{M}_1^3 \left(3 \dot{\psi} + 3 H \phi + 3 \dot{H} \pi - \frac{k^2}{a^2} \pi \right)
  + 2 \bar{M}_2^2 \frac{3 k_0 - k^2}{a^2} \psi
  .\label{eq:poisson}
\end{align}
Similarly, the modified anisotropic shear stress equation is given by
\begin{align}
  m_0^2 \Omega \frac{k^2}{a^2} \left(\psi - \phi\right)
  = P_m \Pi
  + m_0^2 \dot{\Omega} \frac{k^2}{a^2} \pi
  + \bar{M}_2^2 \frac{k^2}{a^2} \left( \Big(H+\frac{2\dot{\bar M}_2}{\bar M_2}\Big)\pi + \phi \right).\label{eq:anisotropic}
\end{align}
Finally, the $\pi$ equation of motion is given by
\begin{align}
  0 ={}&
  c \ddot{\pi} - c \dot{\phi}
  + c \frac{k^2}{a^2} \pi
  + \dot{c} \left(\dot{\pi} - \phi\right)
  + 3 c H \dot{\pi}
  - 3 c \dot{\psi}
  - 6 c H \phi
\nonumber\\&
  + \left( \frac{m_0^2}{4} \dot{\Omega} \dot{R}^0 - 3 c \dot{H} \right) \pi
  - \frac{m_0^2}{2} \dot{\Omega} \left[ -6 (2 H^2 + \dot{H}) \phi - 3 H \dot{\phi} + \frac{k^2}{a^2} \phi - 12 H \dot{\psi} - 3 \ddot{\psi} - 2 \frac{(k^2 - 3k_0)}{a^2} \psi \right]
\nonumber\\&
  + 2 M_2^4\left[ \ddot{\pi} - \dot{\phi} + \left(4 \frac{\dot{M}_2}{M_2} + 3 H\right) (\dot{\pi} - \phi)\right]
\nonumber\\&
  + \frac{\bar{M}_1^3}{2} \Bigg[
  3 \frac{\dot{\bar{M}}_1}{\bar{M}_1} \left(3 \left(\dot{\psi} + H \phi\right) + 3 \dot{H} \pi - \frac{k^2}{a^2} \pi\right)
  + 9 H \left(\dot{\psi} + H \phi\right)
  + 3 (\ddot{H} + 3 H \dot{H}) \pi
  + 3 \ddot{\psi}
\nonumber\\&
  \qquad \qquad + 6 \dot{H} \phi + 3 H \dot{\phi}
  - H \frac{k^2}{a^2} \pi
  - \frac{k^2}{a^2} \phi
  \Bigg]
\nonumber\\&
  + \frac{\bar{M}_2^2}{2}
  \left[6 \left(\dot{H} - \frac{k_0}{a^2} \right) \left(\dot{\psi} + H \phi\right)
  + \left(- 4 \dot{H} \frac{k^2}{a^2} + 6 \dot{H}^2 + 2 \frac{k^2 k_0}{a^4} \right) \pi
  \right]
\nonumber\\&
  + \bar{M}_2^2 \frac{(k^2 - 3 k_0)}{a^2} \left[\left( H + 2 \frac{\dot{\bar{M}}_2}{\bar{M}_2} \right) \psi
  + \left( H^2 + 2 H \frac{\dot{\bar{M}}_2}{\bar{M}_2} + \dot{H} \right) \pi \right]
  .\label{eq:pieom}
\end{align}

\subsection{Quasistatic Approximation}
We now investigate the equations of motion in the limit of the quasistatic approximation, following Refs. \cite{DeFelice2011, Bloomfield2012}. The quasistatic approximation typically consists of two approximations, namely that time derivatives of perturbations are small compared spatial derivatives, and also the sub-horizon approximation $k/aH \gg 1$. We refer the reader to Ref. \cite{Silvestri2013} for a detailed discussion on these approximations.

Under these approximations, the Poisson equation yields
\begin{align}
  \left(2 m_0^2 \Omega
  + 2 \bar{M}_2^2 \right) \psi
  - \left(m_0^2 \dot{\Omega}
  + \bar{M}_1^3\right) \pi
  =
  - \frac{a^2}{k^2} \rho_m \Delta
  \,,
\end{align}
the anisotropic shear stress equation simplifies to
\begin{align}
  - \psi
  + \left(1 + \frac{\bar{M}_2^2}{\Omega m_0^2} \right) \phi
  + \left[\frac{\dot{\Omega}}{\Omega}
  + \frac{\bar{M}_2^2}{\Omega m_0^2} \Big(H+\frac{2\dot{\bar M}_2}{\bar M_2}\Big) \right] \pi = 0\,,
\end{align}
and the $\pi$ equation of motion becomes
\begin{align}
  0 ={}&
  \left[
  m_0^2 \dot{\Omega}
  + \bar{M}_2^2 \left( H + 2 \frac{\dot{\bar{M}}_2}{\bar{M}_2} \right)
  \right] \psi
  + \left[- \frac{m_0^2}{2} \dot{\Omega}
  - \frac{\bar{M}_1^3}{2} \right] \phi
\nonumber\\&
  + \left[c
  - \frac{\bar{M}_1^3}{2} \left(3 \frac{\dot{\bar{M}}_1}{\bar{M}_1} + H\right)
  + \bar{M}_2^2 \left(H^2 - \dot{H} + \frac{k_0}{a^2} + 2 H \frac{\dot{\bar{M}}_2}{\bar{M}_2} \right)
  \right] \pi
\nonumber\\&
  + \left[\frac{m_0^2}{4} \dot{\Omega} \dot{R}^0 - 3 c \dot{H}
  + \frac{3 \bar{M}_1^3}{2} \left(
  3 \dot{H} \frac{\dot{\bar{M}}_1}{\bar{M}_1} + \ddot{H} + 3 H \dot{H}
  \right)
  - 3 \bar{M}_2^2 \frac{k_0}{a^2} \left( H^2 + 2 H \frac{\dot{\bar{M}}_2}{\bar{M}_2} + \dot{H} \right) + 3 \bar{M}_2^2 \dot{H}^2
  \right] \frac{a^2}{k^2} \pi
  \,.
\end{align}
We leave in a mass term for the $\pi$ field, as this may be comparable to $k^2/a^2$. These equations can be combined in a matrix equation,
\begin{align}
  \left(
  \begin{array}{ccc}
    A_\psi & 0 & A_\pi \\
    B_\psi & B_\phi & B_\pi \\
    C_\psi & C_\phi & C_\pi + \frac{a^2}{k^2} C_{\pi2}
  \end{array}
  \right)
  \left(
  \begin{array}{c}
  \psi \\
  \phi \\
  \pi
  \end{array}
  \right)
  =
  \left(
  \begin{array}{c}
  - \frac{a^2}{k^2} \rho_m \Delta \\
  0 \\
  0
  \end{array}
  \right)
\end{align}
where $A_x$ are the coefficients in the Poisson equation, $B_x$ are the coefficients of the anisotropic shear stress equation, and $C_x$ are the coefficients in the $\pi$ equation of motion. Denoting this matrix as ${\cal M}$, the equation can be inverted to obtain
\begin{align}
  \left(
  \begin{array}{c}
  \psi \\
  \phi \\
  \pi
  \end{array}
  \right)
  =
  - \left(
  \begin{array}{c}
  \left[{\cal M}^{-1}\right]_{11} \\
  \left[{\cal M}^{-1}\right]_{12} \\
  \left[{\cal M}^{-1}\right]_{13}
  \end{array}
  \right)
  \frac{a^2}{k^2} \rho_m \Delta \,.
\end{align}

We can now compute the effective Newtonian constant and anisotropic shear stress phenomenological functions, defined by
\begin{align}
  \phi &= - 4 \pi G_{\mathrm{eff}}(k, t) \frac{a^2}{k^2} \rho_m \Delta\,,
\\
  \psi &= \gamma(k, t) \phi \,.
\end{align}
These functions have been given a variety of names by different authors, and have been investigated in a number of papers (see, e.g., \cite{caldwell2007, bean2010, Silvestri2013}). For Horndeski's theory, these functions are given by
\begin{align}
  G_{\mathrm{eff}}(k, t)
  &= \frac{1}{4 \pi} \left[{\cal M}^{-1}\right]_{12}
  = \frac{1}{4 \pi}
  \frac{B_\pi C_\psi - B_\psi C_\pi - B_\psi C_{\pi2} \frac{a^2}{k^2}}
   {A_\psi \left(B_\phi C_\pi + B_\phi C_{\pi2} \frac{a^2}{k^2} - B_\pi C_\phi\right) + A_\pi \left(B_\psi C_\phi - B_\phi C_\psi\right)}\,,
\\
  \gamma(k, t)
  &= \frac{\psi}{\phi}
  = \frac{\left[{\cal M}^{-1}\right]_{11}}{\left[{\cal M}^{-1}\right]_{12}}
  = \frac{B_\phi C_\pi - B_\pi C_\phi + B_\phi C_{\pi 2} \frac{a^2}{k^2}}
  {B_\pi C_\psi - B_\psi C_\pi - B_\psi C_{\pi 2} \frac{a^2}{k^2}}\,.
\end{align}
As stressed by Silvestri \textit{et al.} \cite{Silvestri2013}, the numerator of $G_{\mathrm{eff}}$ and the denominator of $\gamma$ are the same. Both numerator and denominator are polynomials in $a/k$, and as expected for a scalar field theory, only even powers of $a/k$ appear. Furthermore, as expected from Horndeski's theory, only a constant and the second power of $a/k$ are present. Finally, we see that $G_{\mathrm{eff}}$ and $\gamma$ can be written in terms of six functions of time, in the form
\begin{align}
  G_{\mathrm{eff}} = \frac{1}{4 \pi} \frac{f_1 + f_2 \frac{a^2}{k^2}}{f_3 + f_4 \frac{a^2}{k^2}} \,,
\qquad
  \gamma = \frac{f_5 + f_6 \frac{a^2}{k^2}}{f_1 + f_2 \frac{a^2}{k^2}}.
\end{align}
This can be reduced to only five background functions of time by defining $g_i = f_i / f_j$ for a given $j$.

Comparing the function counting, we showed above that the construction of the Horndeski action in the EFT formalism requires six free functions of time. However, note that the function $\Lambda(t)$ does not appear in any of the equations of motion \eqref{eq:poisson}, \eqref{eq:anisotropic} or \eqref{eq:pieom}. Furthermore, in the quasistatic limit, the function $M_2$ does not appear in the matrix ${\cal M}$, and therefore only four free functions are required to specify $G_{\mathrm{eff}}$ and $\gamma$ for Horndeski's theory (assuming $H$ is given).

If the background evolution is completely specified \textit{a priori}, the Friedmann equations \eqref{eq:friedmann1} and \eqref{eq:friedmann2} can (in principle) be used to eliminate two of $\Omega(t)$, $\Lambda(t)$ and $c(t)$ in favour of known background functions of time. Choosing these to be $\Omega(t)$ and $c(t)$, this implies that only two free functions of time are required to specify the behavior of Horndeski's theory in the quasistatic limit. This is a significant reduction in theory space, and bears further investigation. However, in practice, it is easiest to eliminate $c(t)$ and $\Lambda(t)$, leaving three free functions of time.

\section{Conclusions}\label{sec:discussion}

In this paper, we sought to describe linear perturbations in Horndeski's general scalar field theory about an FRW background, using the machinery of the EFT of inflation. We began by motivating the terms in the EFT action that would be required by using a derivative-counting argument, and suggested that only six functions of time would be needed. We then used a perturbative expansion to explicitly match the coefficients of the EFT operators to the free functions in Horndeski's theory.

Once the theory was constructed, we presented the modified Poisson equation, the modified anisotropic shear stress equation, and the scalar field equation of motion. These results provide an independent check of the work of De Felice \textit{et al.} \cite{DeFelice2011}, as well their generalization to include spatial curvature.

We next investigated the behavior of the model in the quasistatic approximation, and used this limit to calculate the effective Newtonian constant and the ratio between the gravitational scalars $\psi/\phi$. It was found that in this limit, only four of the functions of time in the EFT construction appeared in the equations. This presents a reduction beyond the function counting that Silvestri \textit{et al.} achieved from general arguments, although their arguments are more generally applicable.

It was discussed that by fixing the background evolution of the cosmology \textit{a priori}, further reductions in the number of functions of time could be achieved, implying that even fewer functions of time are required to specify the linear behavior of the most general scalar field theory. This motivates investigating how a principle component analysis could best constrain this theory over this reduced parameter space, which is a significant reduction from four functions of two variables.

Although these results have been directed towards dark energy models, the formalism applies equally well to inflationary models (indeed; this is where the formalism was originally applied, with much success). There are two major differences when applying the formalism to inflationary models. Firstly, there is no matter present, and so the matter action can be ignored. Secondly, one can perform a conformal transformation of the metric to set $\Omega(t) \rightarrow 1$. The Friedmann equations then uniquely specify $\Lambda(t)$ and $c(t)$ in terms of $H$. Thus, linear perturbations to Horndeski's theory applied to inflation are described by just four functions of time: $H$, $M_2$, $\bar{M}_1$ and $\bar{M}_2$. This is further reduced if $H(t)$ is specified \textit{a priori}. Unfortunately, inflationary models are typically more interested in extracting the three-point function, which go beyond our present results. Attempting a matching in the action at cubic order is an incredibly daunting task, although we suggest that the terms present at cubic order in the EFT expansion should be able to be determined by using similar arguments to the derivative counting arguments presented here.

We are currently using the EFT approach discussed in this paper to investigate the parameter space of Horndeski's theory, and compare it to observational data, using CAMB \cite{Lewis1999} to numerically evolve the equations of motion. We hope that data from future experiments such as the LSST, the Dark Energy Survey and Euclid will allow us to place stringent constraints on even the most general of theories.

\acknowledgments

We thank Minjoon Park, Eanna Flanagan, Scott Watson, Leo Stein and Rachel Bean for helpful discussions. We thank Jerome Gleyzes, Filippo Vernizzi, Federico Piazza and Giulia Gubitosi for their comments and assistance with cross-checking results. This work was performed using the xTensor \cite{xTensor} and xPand \cite{Pitrou2013} packages for Mathematica. JB was supported by NSF grants PHY-1068541 and PHY-0968820.

Addendum: While this work was in its final stages of preparation, \cite{Gleyzes2013} was published, presenting similar results but using a different analysis. We have checked that all results are in agreement.

\bibliographystyle{apsrev}
\bibliography{darkenergyeft}

\begin{thebibliography}{29}
\expandafter\ifx\csname natexlab\endcsname\relax\def\natexlab#1{#1}\fi
\expandafter\ifx\csname bibnamefont\endcsname\relax
  \def\bibnamefont#1{#1}\fi
\expandafter\ifx\csname bibfnamefont\endcsname\relax
  \def\bibfnamefont#1{#1}\fi
\expandafter\ifx\csname citenamefont\endcsname\relax
  \def\citenamefont#1{#1}\fi
\expandafter\ifx\csname url\endcsname\relax
  \def\url#1{\texttt{#1}}\fi
\expandafter\ifx\csname urlprefix\endcsname\relax\def\urlprefix{URL }\fi
\providecommand{\bibinfo}[2]{#2}
\providecommand{\eprint}[2][]{\url{#2}}

\bibitem[{\citenamefont{Riess et~al.}(1998)}]{Riess1998}
\bibinfo{author}{\bibfnamefont{A.~G.} \bibnamefont{Riess}} \bibnamefont{et~al.}
  (\bibinfo{collaboration}{Supernova Search Team}), \bibinfo{journal}{The
  Astronomical Journal} \textbf{\bibinfo{volume}{116}}, \bibinfo{pages}{1009}
  (\bibinfo{year}{1998}), \eprint{astro-ph/9805201}.

\bibitem[{\citenamefont{Perlmutter et~al.}(1999)}]{Perlmutter1999}
\bibinfo{author}{\bibfnamefont{S.}~\bibnamefont{Perlmutter}}
  \bibnamefont{et~al.} (\bibinfo{collaboration}{Supernova Cosmology Project}),
  \bibinfo{journal}{The Astrophysical Journal} \textbf{\bibinfo{volume}{517}},
  \bibinfo{pages}{565} (\bibinfo{year}{1999}), \eprint{astro-ph/9812133}.

\bibitem[{\citenamefont{Clifton et~al.}(2012)\citenamefont{Clifton, Ferreira,
  Padilla, and Skordis}}]{Skordis2011}
\bibinfo{author}{\bibfnamefont{T.}~\bibnamefont{Clifton}},
  \bibinfo{author}{\bibfnamefont{P.~G.} \bibnamefont{Ferreira}},
  \bibinfo{author}{\bibfnamefont{A.}~\bibnamefont{Padilla}}, \bibnamefont{and}
  \bibinfo{author}{\bibfnamefont{C.}~\bibnamefont{Skordis}},
  \bibinfo{journal}{Physics Reports} \textbf{\bibinfo{volume}{513}},
  \bibinfo{pages}{1} (\bibinfo{year}{2012}), \eprint{1106.2476}.

\bibitem[{\citenamefont{Huterer and Turner}(1999)}]{Huterer:1998qv}
\bibinfo{author}{\bibfnamefont{D.}~\bibnamefont{Huterer}} \bibnamefont{and}
  \bibinfo{author}{\bibfnamefont{M.~S.} \bibnamefont{Turner}},
  \bibinfo{journal}{Phys.Rev.} \textbf{\bibinfo{volume}{D60}},
  \bibinfo{pages}{081301} (\bibinfo{year}{1999}), \eprint{astro-ph/9808133}.

\bibitem[{\citenamefont{Starobinsky}(1998)}]{Starobinsky:1998fr}
\bibinfo{author}{\bibfnamefont{A.~A.} \bibnamefont{Starobinsky}},
  \bibinfo{journal}{JETP Letters} \textbf{\bibinfo{volume}{68}},
  \bibinfo{pages}{757} (\bibinfo{year}{1998}), \eprint{astro-ph/9810431}.

\bibitem[{\citenamefont{Nakamura and Chiba}(1999)}]{Nakamura:1998mt}
\bibinfo{author}{\bibfnamefont{T.}~\bibnamefont{Nakamura}} \bibnamefont{and}
  \bibinfo{author}{\bibfnamefont{T.}~\bibnamefont{Chiba}},
  \bibinfo{journal}{Mon.Not.Roy.Astron.Soc.} \textbf{\bibinfo{volume}{306}},
  \bibinfo{pages}{696} (\bibinfo{year}{1999}), \eprint{astro-ph/9810447}.

\bibitem[{\citenamefont{Boisseau et~al.}(2000)\citenamefont{Boisseau,
  Esposito-Farese, Polarski, and Starobinsky}}]{Boisseau:2000pr}
\bibinfo{author}{\bibfnamefont{B.}~\bibnamefont{Boisseau}},
  \bibinfo{author}{\bibfnamefont{G.}~\bibnamefont{Esposito-Farese}},
  \bibinfo{author}{\bibfnamefont{D.}~\bibnamefont{Polarski}}, \bibnamefont{and}
  \bibinfo{author}{\bibfnamefont{A.~A.} \bibnamefont{Starobinsky}},
  \bibinfo{journal}{Phys.Rev.Lett.} \textbf{\bibinfo{volume}{85}},
  \bibinfo{pages}{2236} (\bibinfo{year}{2000}), \eprint{gr-qc/0001066}.

\bibitem[{\citenamefont{Park et~al.}(2010)\citenamefont{Park, Zurek, and
  Watson}}]{Park2010}
\bibinfo{author}{\bibfnamefont{M.}~\bibnamefont{Park}},
  \bibinfo{author}{\bibfnamefont{K.~M.} \bibnamefont{Zurek}}, \bibnamefont{and}
  \bibinfo{author}{\bibfnamefont{S.}~\bibnamefont{Watson}},
  \bibinfo{journal}{Physical Review D} \textbf{\bibinfo{volume}{81}},
  \bibinfo{pages}{124008} (\bibinfo{year}{2010}), \eprint{1003.1722}.

\bibitem[{\citenamefont{Bloomfield and Flanagan}(2012)}]{Bloomfield:2011wa}
\bibinfo{author}{\bibfnamefont{J.~K.} \bibnamefont{Bloomfield}}
  \bibnamefont{and} \bibinfo{author}{\bibfnamefont{E.~E.}
  \bibnamefont{Flanagan}}, \bibinfo{journal}{Journal of Cosmology and
  Astroparticle Physics} \textbf{\bibinfo{volume}{2012}}, \bibinfo{pages}{039}
  (\bibinfo{year}{2012}), \eprint{1112.0303}.

\bibitem[{\citenamefont{Gubitosi et~al.}(2013)\citenamefont{Gubitosi, Piazza,
  and Vernizzi}}]{Gubitosi2012}
\bibinfo{author}{\bibfnamefont{G.}~\bibnamefont{Gubitosi}},
  \bibinfo{author}{\bibfnamefont{F.}~\bibnamefont{Piazza}}, \bibnamefont{and}
  \bibinfo{author}{\bibfnamefont{F.}~\bibnamefont{Vernizzi}},
  \bibinfo{journal}{JCAP} \textbf{\bibinfo{volume}{1302}}, \bibinfo{pages}{032}
  (\bibinfo{year}{2013}), \eprint{1210.0201}.

\bibitem[{\citenamefont{Bloomfield et~al.}(2012)\citenamefont{Bloomfield,
  Flanagan, Park, and Watson}}]{Bloomfield2012}
\bibinfo{author}{\bibfnamefont{J.~K.} \bibnamefont{Bloomfield}},
  \bibinfo{author}{\bibfnamefont{E.~E.} \bibnamefont{Flanagan}},
  \bibinfo{author}{\bibfnamefont{M.}~\bibnamefont{Park}}, \bibnamefont{and}
  \bibinfo{author}{\bibfnamefont{S.}~\bibnamefont{Watson}}
  (\bibinfo{year}{2012}), \eprint{1211.7054}.

\bibitem[{\citenamefont{Baker et~al.}(2013)\citenamefont{Baker, Ferreira, and
  Skordis}}]{Baker2012}
\bibinfo{author}{\bibfnamefont{T.}~\bibnamefont{Baker}},
  \bibinfo{author}{\bibfnamefont{P.~G.} \bibnamefont{Ferreira}},
  \bibnamefont{and} \bibinfo{author}{\bibfnamefont{C.}~\bibnamefont{Skordis}},
  \bibinfo{journal}{Phys.Rev.} \textbf{\bibinfo{volume}{D87}},
  \bibinfo{pages}{024015} (\bibinfo{year}{2013}), \eprint{1209.2117}.

\bibitem[{\citenamefont{Silvestri et~al.}(2013)\citenamefont{Silvestri,
  Pogosian, and Buniy}}]{Silvestri2013}
\bibinfo{author}{\bibfnamefont{A.}~\bibnamefont{Silvestri}},
  \bibinfo{author}{\bibfnamefont{L.}~\bibnamefont{Pogosian}}, \bibnamefont{and}
  \bibinfo{author}{\bibfnamefont{R.~V.} \bibnamefont{Buniy}}
  (\bibinfo{year}{2013}), \eprint{1302.1193}.

\bibitem[{\citenamefont{Charmousis et~al.}(2012)\citenamefont{Charmousis,
  Copeland, Padilla, and Saffin}}]{Charmousis2011}
\bibinfo{author}{\bibfnamefont{C.}~\bibnamefont{Charmousis}},
  \bibinfo{author}{\bibfnamefont{E.~J.} \bibnamefont{Copeland}},
  \bibinfo{author}{\bibfnamefont{A.}~\bibnamefont{Padilla}}, \bibnamefont{and}
  \bibinfo{author}{\bibfnamefont{P.~M.} \bibnamefont{Saffin}},
  \bibinfo{journal}{Physical Review Letters} \textbf{\bibinfo{volume}{108}},
  \bibinfo{pages}{051101} (\bibinfo{year}{2012}), \eprint{1106.2000}.

\bibitem[{\citenamefont{Battye and Pearson}(2012)}]{battye2012}
\bibinfo{author}{\bibfnamefont{R.~A.} \bibnamefont{Battye}} \bibnamefont{and}
  \bibinfo{author}{\bibfnamefont{J.~A.} \bibnamefont{Pearson}},
  \bibinfo{journal}{Journal of Cosmology and Astroparticle Physics}
  \textbf{\bibinfo{volume}{1207}}, \bibinfo{pages}{019} (\bibinfo{year}{2012}),
  \eprint{1203.0398}.

\bibitem[{\citenamefont{Horndeski}(1974)}]{Horndeski1974}
\bibinfo{author}{\bibfnamefont{G.~W.} \bibnamefont{Horndeski}},
  \bibinfo{journal}{International Journal of Theoretical Physics}
  \textbf{\bibinfo{volume}{10}}, \bibinfo{pages}{363} (\bibinfo{year}{1974}),
  ISSN \bibinfo{issn}{0020-7748}.

\bibitem[{\citenamefont{Deffayet et~al.}(2011)\citenamefont{Deffayet, Gao,
  Steer, and Zahariade}}]{Deffayet2011}
\bibinfo{author}{\bibfnamefont{C.}~\bibnamefont{Deffayet}},
  \bibinfo{author}{\bibfnamefont{X.}~\bibnamefont{Gao}},
  \bibinfo{author}{\bibfnamefont{D.}~\bibnamefont{Steer}}, \bibnamefont{and}
  \bibinfo{author}{\bibfnamefont{G.}~\bibnamefont{Zahariade}},
  \bibinfo{journal}{Physical Review D} \textbf{\bibinfo{volume}{84}},
  \bibinfo{pages}{064039} (\bibinfo{year}{2011}), \eprint{1103.3260}.

\bibitem[{\citenamefont{Kobayashi et~al.}(2011)\citenamefont{Kobayashi,
  Yamaguchi, and Yokoyama}}]{Kobayashi2011}
\bibinfo{author}{\bibfnamefont{T.}~\bibnamefont{Kobayashi}},
  \bibinfo{author}{\bibfnamefont{M.}~\bibnamefont{Yamaguchi}},
  \bibnamefont{and} \bibinfo{author}{\bibfnamefont{J.}~\bibnamefont{Yokoyama}},
  \bibinfo{journal}{Progress of Theoretical Physics}
  \textbf{\bibinfo{volume}{126}}, \bibinfo{pages}{511} (\bibinfo{year}{2011}),
  \eprint{1105.5723}.

\bibitem[{\citenamefont{Horava}(2009)}]{Horava2009}
\bibinfo{author}{\bibfnamefont{P.}~\bibnamefont{Horava}},
  \bibinfo{journal}{Phys.Rev.} \textbf{\bibinfo{volume}{D79}},
  \bibinfo{pages}{084008} (\bibinfo{year}{2009}), \eprint{0901.3775}.

\bibitem[{\citenamefont{Cheung et~al.}(2008)\citenamefont{Cheung, Creminelli,
  Fitzpatrick, Kaplan, and Senatore}}]{Cheung2008}
\bibinfo{author}{\bibfnamefont{C.}~\bibnamefont{Cheung}},
  \bibinfo{author}{\bibfnamefont{P.}~\bibnamefont{Creminelli}},
  \bibinfo{author}{\bibfnamefont{A.~L.} \bibnamefont{Fitzpatrick}},
  \bibinfo{author}{\bibfnamefont{J.}~\bibnamefont{Kaplan}}, \bibnamefont{and}
  \bibinfo{author}{\bibfnamefont{L.}~\bibnamefont{Senatore}},
  \bibinfo{journal}{Journal of High Energy Physics}
  \textbf{\bibinfo{volume}{0803}}, \bibinfo{pages}{014} (\bibinfo{year}{2008}),
  \eprint{0709.0293}.

\bibitem[{\citenamefont{Creminelli et~al.}(2009)\citenamefont{Creminelli,
  D'Amico, Norena, and Vernizzi}}]{Creminelli2009}
\bibinfo{author}{\bibfnamefont{P.}~\bibnamefont{Creminelli}},
  \bibinfo{author}{\bibfnamefont{G.}~\bibnamefont{D'Amico}},
  \bibinfo{author}{\bibfnamefont{J.}~\bibnamefont{Norena}}, \bibnamefont{and}
  \bibinfo{author}{\bibfnamefont{F.}~\bibnamefont{Vernizzi}},
  \bibinfo{journal}{Journal of Cosmology and Astroparticle Physics}
  \textbf{\bibinfo{volume}{0902}}, \bibinfo{pages}{018} (\bibinfo{year}{2009}),
  \eprint{0811.0827}.

\bibitem[{\citenamefont{De~Felice et~al.}(2011)\citenamefont{De~Felice,
  Kobayashi, and Tsujikawa}}]{DeFelice2011}
\bibinfo{author}{\bibfnamefont{A.}~\bibnamefont{De~Felice}},
  \bibinfo{author}{\bibfnamefont{T.}~\bibnamefont{Kobayashi}},
  \bibnamefont{and}
  \bibinfo{author}{\bibfnamefont{S.}~\bibnamefont{Tsujikawa}},
  \bibinfo{journal}{Physics Letters B} \textbf{\bibinfo{volume}{706}},
  \bibinfo{pages}{123} (\bibinfo{year}{2011}), \eprint{1108.4242}.

\bibitem[{\citenamefont{Gleyzes et~al.}(2013)\citenamefont{Gleyzes, Langlois,
  Piazza, and Vernizzi}}]{Gleyzes2013}
\bibinfo{author}{\bibfnamefont{J.}~\bibnamefont{Gleyzes}},
  \bibinfo{author}{\bibfnamefont{D.}~\bibnamefont{Langlois}},
  \bibinfo{author}{\bibfnamefont{F.}~\bibnamefont{Piazza}}, \bibnamefont{and}
  \bibinfo{author}{\bibfnamefont{F.}~\bibnamefont{Vernizzi}},
  \bibinfo{journal}{JCAP} \textbf{\bibinfo{volume}{1308}}, \bibinfo{pages}{025}
  (\bibinfo{year}{2013}), \eprint{1304.4840}.

\bibitem[{\citenamefont{Kodama and Sasaki}(1984)}]{kodama1984}
\bibinfo{author}{\bibfnamefont{H.}~\bibnamefont{Kodama}} \bibnamefont{and}
  \bibinfo{author}{\bibfnamefont{M.}~\bibnamefont{Sasaki}},
  \bibinfo{journal}{Progress of Theoretical Physics. Supplement}
  \textbf{\bibinfo{volume}{78}}, \bibinfo{pages}{1} (\bibinfo{year}{1984}).

\bibitem[{\citenamefont{Caldwell et~al.}(2007)\citenamefont{Caldwell, Cooray,
  and Melchiorri}}]{caldwell2007}
\bibinfo{author}{\bibfnamefont{R.}~\bibnamefont{Caldwell}},
  \bibinfo{author}{\bibfnamefont{A.}~\bibnamefont{Cooray}}, \bibnamefont{and}
  \bibinfo{author}{\bibfnamefont{A.}~\bibnamefont{Melchiorri}},
  \bibinfo{journal}{Physical Review D} \textbf{\bibinfo{volume}{76}},
  \bibinfo{pages}{023507} (\bibinfo{year}{2007}), \eprint{astro-ph/0703375}.

\bibitem[{\citenamefont{Bean and Tangmatitham}(2010)}]{bean2010}
\bibinfo{author}{\bibfnamefont{R.}~\bibnamefont{Bean}} \bibnamefont{and}
  \bibinfo{author}{\bibfnamefont{M.}~\bibnamefont{Tangmatitham}},
  \bibinfo{journal}{Physical Review D} \textbf{\bibinfo{volume}{81}},
  \bibinfo{pages}{083534} (\bibinfo{year}{2010}), \eprint{1002.4197}.

\bibitem[{\citenamefont{Lewis et~al.}(2000)\citenamefont{Lewis, Challinor, and
  Lasenby}}]{Lewis1999}
\bibinfo{author}{\bibfnamefont{A.}~\bibnamefont{Lewis}},
  \bibinfo{author}{\bibfnamefont{A.}~\bibnamefont{Challinor}},
  \bibnamefont{and} \bibinfo{author}{\bibfnamefont{A.}~\bibnamefont{Lasenby}},
  \bibinfo{journal}{The Astrophysical Journal} \textbf{\bibinfo{volume}{538}},
  \bibinfo{pages}{473} (\bibinfo{year}{2000}), \eprint{astro-ph/9911177}.

\bibitem[{\citenamefont{Martin-Garcia}()}]{xTensor}
\bibinfo{author}{\bibfnamefont{J.~M.} \bibnamefont{Martin-Garcia}},
  \emph{\bibinfo{title}{x{T}ensor}}, \urlprefix\url{http://www.xact.es/}.

\bibitem[{\citenamefont{Pitrou et~al.}(2013)\citenamefont{Pitrou, Roy, and
  Umeh}}]{Pitrou2013}
\bibinfo{author}{\bibfnamefont{C.}~\bibnamefont{Pitrou}},
  \bibinfo{author}{\bibfnamefont{X.}~\bibnamefont{Roy}}, \bibnamefont{and}
  \bibinfo{author}{\bibfnamefont{O.}~\bibnamefont{Umeh}}
  (\bibinfo{year}{2013}), \eprint{1302.6174},
  \urlprefix\url{http://www2.iap.fr/users/pitrou/xpand.htm}.

\end{thebibliography}

\end{document}